-
\documentstyle[preprint,aps,eqsecnum]{revtex}

\begin{document}

\def\ltap{\;\raisebox{-.4ex}{\rlap{$\sim$}} \raisebox{.4ex}{$<$}\;}
\def\gtap{\;\raisebox{-.4ex}{\rlap{$\sim$}} \raisebox{.4ex}{$>$}\;}

{\tighten
\preprint{\vbox{\hbox{FERMILAB--PUB--94--163--T}\hbox{\today}}}
\title{Distinguishing $B$ and $\overline B$ Hadrons}

\author{Isard Dunietz}
\address{Fermi National Accelerator Laboratory \\ P.O.~Box 500,
Batavia, IL 60510}


\maketitle
\begin {abstract}
Distinguishing the flavor of $B$ and $\overline B$ hadrons is
critical
in studies of CP-violation, $B^0 -\overline{B^0}$ mixing, and the
underlying $b$-decay mechanisms. Methods of $b$ ``flavor tagging"
are broadly divided into ``opposite $b$" tagging and self-tagging of
the signal $b$ hadron. The former, while understood,
has the perceived drawback of low efficiency.
The latter, while having the potential for an order of
magnitude higher efficiency, has yet to be demonstrated for neutral
$B$
hadrons. In this article we review opposite $b$ tagging in light of
methods whose efficacy has only recently been demonstrated or
suggested.
In addition, we recommend a number of tagging methods for the
opposite $b$
including:
$K^{*0}$ and $K^{*\pm}$ with large inclusive yields of 15\% and 18\%;
$\overline \Lambda$ and $\overline {\Lambda}p $; partially
reconstructed charmed hadrons; sophisticated jet charge
techniques, etc. We also recommend the use of self-tagging
for the $opposite$ $b$ hadron. Such an inversion of self-tagging
could conceivably increase the efficiency of opposite $b$ tagging and
even mitigate the effects of neutral $B$ mixing.
Self-tagging of the signal hadron, when possible, could be used
either by itself or to
confirm the result from the opposite $b$ tag. We suggest that all
methods be weighted by their dilutions and combined to
yield efficient tagging. For a given detector,
this requires that the dilution of each tag be well measured. We
therefore
review the determination of the dilution $D_T$ for a general tag $T$
in some detail. Finally, we briefly consider CP-violation in the $b$
sector
and suggest a number of exclusive modes which can be combined for
higher
statistics probes of the unitarity angles. While ambitious from an
experimental perspective, the program of flavor identification
outlined here
has the potential to yield important fundamental results in the near
future.

\end{abstract}

}

\section{Introduction}

Identifying (``tagging") the flavor of beauty hadrons is crucial
in studies of CP violation, $B_s -\overline {B_s}$ mixing, and
in separating inclusive or exclusive yields of $b$-hadrons
vs.~$\overline b$-hadrons.
This note advocates to weight and combine all conceivable tags in
order
to optimize the flavor identification of neutral $B$ mesons.
For such a program to succeed, the purity (henceforth referred to as
dilution)
of each tag must be determined.  Fortunately, for any
experimental setup the dilution for each tag $T$ can be
determined from $B^\pm T$, primary $\ell^\pm T$,
and/or $\stackrel{(-)}{B_d} T$ correlations.  Note that we denote
as ``primary" the lepton coming directly from the decay of the $b$.
These correlations
could initially be compared for consistency and, once they are
understood, they could be combined to yield a higher statistics
measure of
dilution for each tag.  Studies
of $b$-decay dynamics, of $B_s -\overline{B_s}$ mixing, and of CP
violation then become feasible by pairing the relevant mode under
study
with any available tag $T$.
Consider for instance the CP violating asymmetry of $B_d\rightarrow
J/\psi K_S$.
Having previously measured dilutions
we can correct the asymmetry measurement for the impurities of
each tagging method. The sum of available tags $T$ optimizes the
measurement
of the undiluted asymmetry.

A number of schemes for distinguishing $B^0$ and $\overline
{B}^0$ mesons have been demonstrated or suggested. These include:
(i) self-tagging\cite{gnr,gr,ab,wicklund},
(ii) tagging the flavor of the other $b$ in the event
\cite{leptontag,ktag,lambdatag,dr},
(iii) jet charge, $Q_j$, tagging  \cite{sharma,jetcharge}, and
(iv) polarization-tagging \cite{adg}.
This note explores the first three of these methods.
The first identifies the initial
flavor of the $B$ meson from the charge of a primary
fragmentation hadron produced
nearby in
phase-space.  Throughout this note, we denote a primary hadron as one
that originates from the
primary interaction vertex, while a primary lepton is a lepton from
$b$-decay and is normally displaced from (i.e. has significant impact
parameter with respect to) the primary vertex. Self-tagging
suggests that $K^+ B_s (K^-\overline B_s)$ events could be enhanced
over $K^-B_s (K^+ \overline{B_s})$ events~\cite{ab}. Similarly, there
could be more $K^* B_s (\overline K^* \overline B_s)$ than
$\overline K^* B_s (K^* \overline{B_s})$ events \cite{wicklund},
where $K^*$ is either neutral or charged.   Ref. \cite{gnr}
predicts an enhancement of
$\pi^+ B^- (\pi^- B^+)$ over $\pi^-B^-(\pi^+B^+)$ events, and of
$\pi^+ B_d
(\pi^-\overline B_d)$ over
$\pi^- B_d(\pi^+ \overline{B}_d)$ events.
The CDF collaboration is in the midst of studying the feasibility of
self-tagging \cite{shochet}.  If self-tagging works for neutral $B$'s
it could be employed directly to tag the signal $B$. Even if it does
not, it may still work for charged $B$'s.
In either case, self-tagging can ``literally" be inverted and used
effectively
to tag the flavor of the opposite $b$ hadron as will be discussed
below.

Whereas self-tagging may or may not work, tagging the flavor
of the other $b$-hadron in the event \underline{must} work in
principle.
Consider a $b\overline b$
event where the $\overline b$ hadronizes into a $B$ which is seen in
a decay-mode under study, such as
$B_d\rightarrow J/\psi K_S,\; \pi^+\pi^-,\; B_s \rightarrow J/\psi
\phi , \;
\; D^{-}_{s} l\nu ,\; D^\pm_s K^\mp,\; D^{-}_{s}\pi^+$, etc. The
other $b$ hadronizes into any of many beautiful species $\overline
B_d,\;
B^-,\; \overline B_s, \;\Lambda_b,\; \Xi_b$, etc.
The relative mix of beauty hadrons depends upon how the $b\overline
b$ pair
is produced. The production fractions are roughly

\begin{equation}
\overline{B_d} \;\text{:}\; B^-\approx 0.5 \;\text{:}\; 0.5
\end{equation}
for an $e^+ e^- \to \Upsilon(4S)\to B\overline{B}$ experiment.
And they are

\begin{equation}
\overline{B_d} \;\text{:}\; B^- \;\text{:}\; \overline{B_s}
\;\text{:}\;\Lambda_b
\approx 0.375 \;\text{:}\; 0.375 \;\text{:}\; 0.15 \;\text{:}\;0.10
\end{equation}
for a high energy experiment,
$e^+e^- \to Z^0 \to b\overline b$ or $p\overline p \to b\overline b
+~X\ldots \;.$

For a given high energy experiment, we denote the mix of
the various beautiful species by $H_b$.  {\it $H_b$
\underline{rarely}
loses $b$-flavor information due to $B^0 -\overline{B^0}$ mixing.}
This can be quantified in terms of a dilution parameter $D$,
\begin{equation}
D\equiv \frac{\text{Prob} (H_{b,phys}\to H_b ) - \text{Prob}
(H_{b,phys}
\to \overline{H}_b )}
{\text{Prob} (H_{b,phys}\to H_b ) +\text{Prob} (H_{b,phys}\to
\overline{H}_b)} \;.
\end{equation}
Here $H_{b,phys}$ denotes a time-evolved, initially pure $b$-flavored
hadron $H_b$, and $\overline{H}_{b,phys}$ is defined
analogously.

The probability of an initial $H_b$ to oscillate into its
antiparticle is
\begin{eqnarray}
\text{Prob} (H_{b,phys}\to \overline{H}_b) & = & f_d \;
\text{Prob}(\overline{B}_{d,phys}\to B_d ) \nonumber \\
& + & f_s \;\text{Prob}(\overline{B}_{s,phys}\to B_s )
\nonumber \\
& \approx & 0.375 \cdot 0.16 + 0.15 \cdot 0.5 \approx 0.13 \;.
\end{eqnarray}
The production fractions of the various $b$-species are denoted
by $f_u , f_d , f_s , f_{\Lambda_b}$ for $B^- ,\overline{B}_d ,
\overline{B}_s, $ and $\Lambda_b$.  Eq.~(1.4) uses the
known $B_d -\overline{B_d}$ mixing parameter, and assumes
maximal mixing for the $B_s$ meson \cite{sharma,schroder}.
In addition to a respectable---i.e., large---dilution of
\begin{equation}
D\approx 0.74 \;,
\end{equation}
we stress that almost all decays of $H_b$ are flavor-specific---that
is,
an ``ideal detector" is able to flavor-tag nearly every decay of
$H_b$.  The CKM-favored transitions, $b\to c\overline u d,
c\ell\overline{\nu},
c\overline c s$, are generally seen in flavor-specific final states,
with only a few exceptions.  The first, which has minor effect, is
that of
CKM-favored transitions of the $B^0$ which give rise to $K_S$ or
$K_L$
final states in which the original $b$-flavor is lost. In the second,
the $b\to
c\overline{c}s$ transition of the $B_s$ is not flavor-specific, but
this
has no effect on the above derivation of $D$ because maximal $B_s
-\overline{B_s}$ mixing was assumed.

We remark parenthetically that an ideal detector could study
the time-evolution of $B^0$ modes. Thus, since
$B_d -\overline{B_d}$ mixing is known, and $B_s -\overline{B_s}$
mixing will likely be measured in the future, the time-evolution of
the
flavor-specific modes  of the neutral $B^0$ can be partially
disentangled to yield a dilution nearer to unity.
It may even be possible to extract partial flavor information
from the $b\to c\overline{c}s$ transition of the $B_s$
if it so happens, for example, that the $B_s$ prefers
$D^{*+}_s D^-_s$ to $D^+_s  D^{*-}_s$ or vice versa.

By setting a substantial lower limit
on $B_s -\overline{B_s}$ mixing, $(\Delta m/\Gamma)_{B_s}
\;\raisebox{-.4ex}{\rlap{$\sim$}} \raisebox{.4ex}{$>$}\; 9$
\cite{sharma},
it has been recently demonstrated at LEP
that the jet charge, $Q_j$, technique is
a formidable tagging tool \cite{jetcharge}.  The jet charge $Q_j$ is
a
kinematically weighted average of the charges of particles in each
jet.
It uses not only the opposite $b$-jet in the event, but also the
signal jet.  (Naively, a $b$-jet starts with a
charge of $- 1/3$ and is thus more likely to be negatively charged.)
An $e^+e^- \to Z^0 \to b\overline b$  experiment deals with well
balanced $b$-jets, ideally suited for measuring $Q_j$.  In contrast,
at hadron accelerators $b$-jets are not always well separated (as in
the case
$g~\to~b\overline{b}$). In both environments silicon vertex
detectors now afford significant additional guidance via the
detection and
characterization of displaced vertices.  The
requirement of displaced vertices in jets reduces the mistagging rate
of $Q_j$.
To fully optimize jet-charge tagging one should weight displaced
tracks differently from the tracks associated to the primary
interaction
vertex\cite{ikss}.  Further enhancement can be achieved by using any
other
available discriminating information such as the probability that a
given track corresponds to a specific particle type \cite{ikss}.
A sophisticated jet charge algorithm of this type is now
under investigation and could become a powerful tagging
tool~\cite{dikss}.

In contrast to high energy experiments, one could consider
a threshold machine $e^+e^- \to \Upsilon (4S)$. The $\Upsilon (4S)$
is seen in the two-body, p-wave modes; $B^{+}B^{-}$ and $B_d
\overline{B_d}.$
Bose-Einstein statistics forbids a simultaneous
$B^0B^0 (\overline{B^0}\;\overline{B^0})$ state. Thus, the time
of a flavor-specific decay of a neutral $B$ starts the clock for
the time evolution of its partner. Time-dependent measurements allow
studies of CP violation and are one of the main motivations for
asymmetric
$B$-factories at the $\Upsilon (4S)$ \cite{oddone}.  Also at the
$\Upsilon (4S)$, tagging the $B_d$ via the flavor of its partner
meson works very well in principle
\cite{kaleksan}.

Throughout this note, charge conjugate modes and correlations
are implicit, except for the case of CP violation.
Extending these ideas to non charge-symmetric initial states, such as
$pp$ colliders or fixed target options, is straightforward but the
algebra and measurements are more involved and are not discussed
here.
We do not expect any observable coherence
effects at high energy machines, and assume henceforth
incoherent $b\overline b$ production \cite{incoherence}.
Since any detector is imperfect, we have to make do with
incomplete information. We therefore advocate to weight and properly
combine
all conceivable tags.  Although the main thrust of this paper is
to employ the other $b$-hadron as a tag, we also
consider tags that originate from the signal $b$-jet itself, such as
self-tagging and sophisticated jet charge algorithms.  Such a program
truly optimizes tagging.

It is important to distinguish between $B^0$-tagging and
tagging-calibrations in which flavor on one side is known (perhaps
with some
impurity) and flavor tagging efficiency on the other side is being
measured.
The calibrations are essential to the success of this program as they
determine the dilution $D_T$ for each tag and thus allow the correct
weighting of individual tags.  The determination of dilution
for each tag $T$ can be accomplished via several methods, which can
serve first as cross-checks and later be combined to yield a more
accurate measurement of $D_T$.  The dilution for any tag $T$ can be
obtained, for instance, from primary $\ell^\pm T$, flavor specific
$\stackrel{(-)}{B_d} T$, and  $B^\pm T$ correlations.

We discuss in some detail possible data samples of charged $B$'s,
with which to measure $B^\pm T$ correlations.
Whereas current $B^\pm$ data samples include fully reconstructed
$B^\pm \to J/\psi K^\pm$ events, we suggest the use of the much
larger
$B^\pm \to J/\psi X^\pm$ sample where $X^\pm$ denotes an odd
number of charged tracks associated to the $J/\psi$ vertex. The
latter
can be required to be significantly displaced from the primary
interaction
vertex to guarantee $b$-parentage.
Identification of the particles associated to the $J/\psi$ vertex,
while helpful, is not strictly necessary since only the
charges of the particles are relevant. Missing neutrals---such as
$K_S, K_L ,
\pi^0$  or $\gamma$--- contribute no net charge and hence
pose no problem in collecting a $B^\pm$ data sample.
More than a quarter of the inclusive $J/\psi$ yield in $B^\pm$ decays
involve a single charge \cite{psiglasgow,browder},
$$B(B^\pm \to J/\psi K^\pm )=(0.110\; \pm \; 0.015\; \pm \; 0.009)\%
\;,$$
$$B(B^\pm \to J/\psi K^{*\pm}) =(0.178\; \pm \; 0.051\; \pm \;
0.023)\% \;,$$
$$B(B\to J/\psi X) =(1.11\; \pm \; 0.08)\% \;.$$
One could enhance any $B^\pm$ data sample by requiring an
oppositely charged primary hadron nearby in phase space, if
self-tagging works for charged $B$'s~\cite{gnr}.

Ideas exist for collecting even more inclusive data samples of
$b$-hadrons\cite{ikss}. One could for example use $D^{(*)} \ell^- X$
events
that are consistent with
$b$-decays or even combine tracks from secondary and tertiary
vertices to
calculate a vertex mass $m'$ that could distinguish decays
of the heaviest charmed hadrons from those of $b$ hadrons
by exploiting the expectation that in many cases
$$m'_{\Xi_c^0}\;\;(m'_{\Omega_c})\;\;
\;\raisebox{-.4ex}{\rlap{$\sim$}}
\raisebox{.4ex}{$<$}\;
m'_{H_b} \;.$$
The $\Omega_c$ appears in parenthesis because its extremely short
lifetime helps distinguish it from a much longer lived $b$-hadron.
(For the latter, more speculative ideas the background from collinear
$c\overline c$  production in such data samples
has to be assessed for the case of $p\overline p$ colliders.)

We now resume
our discussion of calibrations and $B^0$-tagging experiments.
Once calibrations are complete, one may
turn to measurements utilizing $B^0$-tagging.  Studies of $B_s
-\overline{B_s}$ mixing could become feasible by pairing all flavor
specific $B_s$ candidates with any conceivable tag $T$.  CDF reports
about a
hundred flavor-specific $B_s$ candidates \cite{Bscdf}.  Studies of CP
violation in $B^0$ decays could also be contemplated.

We cannot overemphasize the importance of good particle
identification (p/K/$\pi$, lepton, etc. separation) and observation
of displaced vertices (including tertiaries in some cases) for
tagging
purposes.  Simple tags based upon such information could be quite
effective.
If charged, displaced kaons
could be well identified, they would provide a powerful
tag, with a very large yield in $b$-decays~\cite{ktag}.  Another
potent tag would
be $K^*$'s, which also have a large inclusive yield in
$b$-decays.  For tagging purposes, it will be advantageous to
determine the strangeness content of partially reconstructed
charmed hadrons in $b$-decays, as we will show below.  Combining good
particle identification and observation of displaced vertices in a
sophisticated jet charge algorithm~\cite{dikss}
would allow a substantial fraction
of all signal $B^0$-events to be tagged with high
purity.  Such a program involves an enormous
experimental effort but offers the potential reward of the
immense riches of $B$ physics and conclusive experimental
tests of theoretical speculations.

This note is organized as follows.  Primary lepton - tag correlations
are the subject of Section II.  They yield the dilution for each tag
$T$ and the ratio of the inclusive yield of $T$ in $b$-decay  versus
$\overline b$-decay, by removing $B^0 -\overline{B^0}$ mixing
effects.  Section III reviews existing lepton - tag correlations from
which many tags can be inferred beyond the traditional
lepton~\cite{leptontag} and charged kaon~\cite{ktag} tags.
Alternative tags are enumerated and reviewed.  An intense
study will seek out and discover many additional and general tags
$T$,
such as
those based upon particular event topologies~\cite{workinggroup}.
The discovery
of new tags could come by observing
strong primary $\ell^\pm T$, flavor specific $\stackrel{(-)}{B_d} T$,
or $B^\pm T$ correlations.   Section III reviews in detail how to
obtain the dilution $D_T$ from $B^\pm T$ correlations and also
discusses
time-dependence of $\stackrel{(-)}{B^0} T$ correlations, which allow
$B_s -\overline{B_s}$ mixing and
various CP violation studies.  Because of
its cardinal import, CP violation is the exclusive topic of Section
IV.  A judicious dilution-weighted combination of all accessible
tags, which
avoids multiple-counting, could make an ambitious $B$ program
feasible.
Section V concludes with a bright outlook.

\section{Lepton-Tag Correlations}

This section considers lepton-tag correlations, where the lepton
is from one $b$ hadron decay and the tag $T$ generally
(but not exclusively) originates from
the \underline{other} $b$-hadron in the event.
At the $\Upsilon (4S)$ one could use a hard lepton (to suppress
the background from secondaries, $b\to c\to \ell^+$) and
angular correlations between $\ell^\pm$ and $T$ to guarantee that the
lepton and tag originate from different $B$ mesons.

A high energy experiment, such as
$e^+e^- \to Z^0 \to b\overline b ,\; p\overline p \to b\overline b
+\ldots \;,$
can use hard, displaced leptons with large transverse momenta,
$P_{T,rel}$,
relative to their jet to suppress backgrounds.
The primary lepton signal is enhanced by pairing it with a
displaced vertex from which a few charged prongs emanate such that
the overall topology is consistent with being a
$b$-hadron~\cite{ikss}.
The tag $T$ could be searched for in the hemisphere opposite to the
lepton to avoid collinear $b\overline b$ and $c\overline c$
backgrounds occurring at hadron colliders.
At least one significant background to this primary lepton sample is
known and removable, namely
$\overline{B}\to DD^-_s X$, where the $D$ provides the wrong sign
lepton and the
$D^-_s$ is responsible for the displaced vertex.

The first part of this section concerns itself with $\Upsilon (4S)$
experiments, where the removal of $B_d -\overline{B_d}$ mixing
effects is discussed. This removal is necessary for extracting the
important quantity,
\begin{equation}
L_T \equiv \frac{B(B\to TX)}{B(\overline{B}\to TX)} \; ,
\end{equation}
which separates the inclusive $T$ yield in $B$ decays into relative
fractions
of $B$ and $\overline{B}$.
This information is crucial for understanding the underlying
$b$-decay
dynamics, as a recent $\overline{B}\to\Lambda_c X$ measurement
amply demonstrated\cite{dcfw,baryonglasgow}.
Previous experimental analyses assumed that the inclusive $\Lambda_c$
yield in $B$ decays is dominated by the $b\to c\overline u d$
transition
\cite{previouslambdac},
whereas a recent note
\cite{dcfw}
suggests that, on the contrary, $b\to c\overline{c}s$ is dominant.
Neither of the two hypotheses could be ruled out with present
data samples \cite{dc107},
except for the recently obtained $\ell^\pm \Lambda_c$ correlations,
which show a $b\to c\overline{u}d$ preference with a significant
$b\to c\overline{c}s$ component \cite{baryonglasgow}.
We therefore advocate that lepton-particle(s) correlations
be measured whenever possible.
An important aspect of $L_T$ is that it identifies good
$b$-flavor tags.

The latter part of this section discusses correlations
at high energy experiments.
We again remove $B^0 -\overline{B^0}$ mixing and
show how to determine the dilution for each tag $T$.  If the tag $T$
consists of decay daughters of the other $b$-hadron in the event,
then the removal of $B^0 -\overline{B^0}$ mixing effects determines
not only the dilution $D_T$ but also the important ratio,
\begin{equation}
I_T \equiv\frac{B(\overline{H}_{b,phys}\to TX)}
{B(H_{b,phys}\to TX)} \;.
\end{equation}
$I_T$ measures the relative yield of $T$ in decays of
time-evolved $H_{b,phys}$ versus $\overline{H}_{b,phys}$.  The
dilution $D_T$ can be determined for any tag $T$, regardless of
whether it
is a primary hadron, a sophisticated jet charge algorithm, or
decay products of the opposite $b$-hadron in the event.  In contrast,
the ratio $I_T$ can be extracted only when the tag $T$ consists of
decay products of the other $b$-hadron in the event.  The
determination of
$I_T$ from $\ell^\pm T$ correlations
incorporates the fact that the two $b$ hadrons mix independently when
produced incoherently at high energy machines.

\subsection{$\Upsilon (4S)$  factory}

Consider an $\Upsilon (4S)$ experiment.
The removal of $B_d -\overline{B}_d$ mixing requires, in addition to
lepton-tag $T$ correlations, measurements of inclusive
branching fractions of the $B^+$ and $B^-$ to $T$.
The latter can be measured for a
sample of events in which one $B$ has been fully
reconstructed \cite{exclusive} or, in the case of an asymmetric $B$
factory,
data in which the charge of one $B$ can be determined even without
full
reconstruction by exploiting the topological separation of
the $B$ and $\overline B$ decays.
The pairing of the $B^\pm$ data sample with tag $T$ from
the
other $B^\mp$ in the event determines
$B(B^- \to TX)$ and $B(B^+ \to TX)$
separately.

Consider next the reconstructed $B^0$ and $\overline{B^0}$ data
samples involving flavor-specific modes\cite{b0id}.
The measurement of $B(B^0
\rightarrow TX)$ and $B(\overline{B^0}\rightarrow TX)$ requires
the removal of $B^0 - \overline{B^0}$ mixing,

\begin{equation}
\frac{N_{B^0 T}}{ N_{B^0}} =(1-p) B(\overline{B^0}\rightarrow
TX) +p \;B(B^0 \rightarrow TX)\,,
\end{equation}

\begin{equation}
\frac{N_{\overline{B^0} T}}{N_{\overline{B^0}}} \,=(1-p) B(
B^0\rightarrow
TX) + p \;B(\overline{B^0} \rightarrow TX)\,.
\end{equation}
Here $p$ is the probability for a time-evolved, initially pure
$B^0$ to be seen as a $\overline{ B^0}$,

\begin{equation}
  p\equiv \text{Prob}(B^{0}_{phys}\to \overline{B^0})\approx
  {x^2\over 2(1+x^2)}\;.
\end{equation}
Where in the last equation ${\Delta\Gamma\over\Gamma}$ has been
neglected based upon
Standard Model estimates, and $x$ is defined as
$x\equiv {\Delta m\over \Gamma}$
\cite{schroder}.
The coherence of the $L=1, \;B^0\overline{ B^0}$ state cancels
possible
interference terms once integrations over
the $B^0$ and $\overline{B^0}$ decay times have been performed,
resulting in Eqs.~(2.3)-(2.4)
\cite{dthesis,strictly}.
Because the fully reconstructed $B$ data sample is rather small, one
should use
\underline
{in addition} the larger $\ell^\pm-T$ sample, which is our next
topic.

Theory predicts equal semileptonic widths for the neutral and charged
$B$'s, but allows for differences in lifetimes and production rates
\cite{widthlifetime}.
The lifetimes and production rates are currently found to be
equivalent
within 20\% experimental
uncertainties\cite{sharma,schroder,cdflifetime},
and are assumed equal for this note. (It is a trivial exercise to
incorporate inequalities once they have been observed.) The
probability $p$ is
obtained directly from the hard, primary dilepton sample,

\begin{equation}
\frac{p}{2} = \frac{N_{\ell^- \ell^-} +N_{\ell^+\ell^+}}
{N_{\ell^- \ell^+} + N_{\ell^- \ell^-} + N_{\ell^+ \ell^+}}
\end{equation}
and has a measured value of
\cite{schroder}

\begin{equation}
{p\over 2}=0.079\pm 0.009 \;.
\end{equation}
Numbers of hard, primary leptons from one $B$ paired with tag $T$
from the other $B$ are,

\begin{eqnarray}
N_{\ell ^+ T} & = & N_{B\overline B}\;{1 \over 2}\;\alpha_\ell
\;\alpha_T \;
 B(B\to
X\ell^+ \nu)\bigg\{B(B^-\to TX)+ \nonumber \\
& + & (1-p)\;B(\overline{B_d}\to TX)+p \; B(B_d\to TX)\bigg\}\;,
\end{eqnarray}

\begin{eqnarray}
N_{\ell ^- T} & = & N_{B\overline B}\;{1 \over 2}\;\alpha_\ell
\;\alpha_T \;
 B(B\to
X\ell^+ \nu)\bigg\{B(B^+\to TX)+ \nonumber \\
& + & (1-p)\;B(B_d\to TX)+p \; B(\overline{B}_d\to TX)\bigg\}\;.
\end{eqnarray}
Here $N_{B\overline{B}}$ denotes the number of $B\overline B$ events,
while
$\alpha_\ell$ and $\alpha_T$ are the detection efficiency
and acceptance factor of hard, primary leptons and $T$,
respectively.
Many of the systematic errors cancel in forming the ratio,

\begin{equation}
\frac{N_{\ell^- T}}{N_{\ell^+ T}}  =  \frac{B(B^+ \to TX) +(1-p)\;
B(B_d\to TX)
+p\;B(\overline{B_d} \to TX)}
{B(B^-\to TX)+(1-p) \;B(\overline{B_d}\to TX)+p\;B(B_d \to TX)} \;.
\end{equation}

Finally, the largest relevant data sample corresponds to inclusive
$T$ in $B$
and $\overline B$ decays,

\begin{equation}
R_T\equiv B(B\to TX)+B(\overline B\to TX),
\end {equation}
where

\begin{equation}
B(B\to TX)\equiv {B(B^+\to TX)+B(B_d\to TX) \over 2}\;,
\end{equation}

\begin{equation}
B(\overline{B} \to TX)\equiv {B(B^-\to TX)+B(\overline{B_d}\to TX)
\over 2}\;.
\end{equation}
By means of this inclusive $T$ sample, the $\ell^\pm T$
sample, the measurement of $p$, and the lower statistics measurements
of the
four separate branching fractions from the fully reconstructed and/or
topologically separated $B$ data sample, it is possible to
both correctly remove $B^0-\overline{ B^0}$ mixing and
to determine the four separate branching fractions, $B^+\to
TX,B^-\to TX, B_d\to TX$ and $\overline{B_d}\to TX$.

In order to remove $B^0-\overline{B^0}$ mixing from existing
$\ell^\pm T$ correlations in the absence of inclusive measurements of
$B(B^\pm \rightarrow TX)$, we \underline{assume} the following
relationships

\begin{equation}
B(B_d \to TX)=B(B^+ \to TX)=B(B \to TX),
\end{equation}

\begin{equation}
B(\overline{B_d}\to TX)=B(B^- \to TX)=B(\overline B \to TX),
\end{equation}

resulting in

\begin{equation}
N_{\ell ^+ T} \sim \left(1-\frac{p}{2} \right)\;
B(\overline B \to TX) + \frac{p}{2} \;
B(B\to TX),
\end{equation}
\begin{equation}
N_{\ell ^- T} \sim \left(1- \frac{p}{2}\right)\;
B(B\to TX) +  \frac{p}{2} \;B(\overline
B \to TX).
\end{equation}
The ratio $L_T$ is determined from $N_{\ell^-
T}/N_{\ell^+ T}$
and the measured value of $p$.
It is a very important quantity, because it probes the
underlying $B$-decay dynamics
and determines how well $T$ tags $b$-flavor.

We stress that the assumptions of Eqs.~(2.14)-(2.15) could very well
be invalid, as the following two extremes illustrate.
The first being the case in which the inclusive yield of tags
$T$ is due to $B^\pm$ decays only. In that case $N_{\ell^-
T}/N_{\ell^+
T}$ determines the ratio $L_T$ without having to correct for any $B_d
-\overline{B_d}$ mixing, since
$$
N_{\ell ^+ T} \sim \;
B(\overline B \to TX) = \; B(B^- \to TX)/2,
$$
$$
N_{\ell ^- T} \sim \;
B(B \to TX) = \; B(B^+ \to TX)/2.
$$
The second extreme is the case in which
the inclusive $T$ yield arises solely from
$\stackrel{(-)}{B_d}$ decays.  In this case
the effect of $B_d -\overline{B_d}$ mixing is maximal and must
be removed to obtain $L_T$, since
$$
N_{\ell ^+ T} \sim \;
(1 - p)\; B(\overline B \to TX) \; +\; p\; B(B \to TX),
$$
$$
N_{\ell ^- T} \sim \;
(1 - p)\; B(B \to TX) \; +\; p\; B(\overline B \to TX).
$$

Separate measurements of $B(B^\pm \to TX)$ are thus crucial for a
correct removal of $B_d -\overline{B_d}$ mixing.  In the absence of
such
$B^\pm \to TX$ measurements, theory can be used as a guide.
For instance, we predict that Eqs.~(2.14)-(2.15) are approximately
true
for tags $T$ such as the sum of charged and neutral $D$'s, or
$D_s$'s.  If no guide is available, we suggest to employ the
``golden mean" which resides halfway between the two extremes and is
precisely Eqs.~(2.14)-(2.17).

For sufficiently large
$\ell^\pm T$ data samples, the understanding of inclusive
$T$ yields may be improved by measuring $L_T$ for various momentum
bins of $T$. It is conceivable that $T$ tagging is enhanced
in particular momentum ranges and observations of such effects
would shed light upon the underlying
$B$-decay mechanism.
The quantity $L_T$ can be used for CP studies at an $\Upsilon (4S)$,
and is a rather reliable barometer of
good $b$-tags at high energy experiments,
barring a few exceptions.

\subsection{High energy experiments}

Lepton-tag correlations at high energy experiments determine
the dilution of each tag $T$. As discussed in Section I, the
tag need not originate from the other $b$-hadron in the event.
But, if it does, then the $\ell^\pm T$ correlations also provide
information
about the
relative fractions of inclusive $T$ production in $b$-hadron versus
$\overline b$-hadron decays,
\begin{equation}
I_T \equiv\frac{B(\overline{H}_{b,phys}\to TX)}
{B(H_{b,phys}\to TX)} \;.
\end{equation}
The inclusive branching fraction of a time-evolved $H_b$ to $T$
is denoted by
\begin{eqnarray}
B(H_{b,phys}\to TX) & = &  f_u \;B(B^-_u \to TX) + \nonumber \\
& + & f_d \;B(\overline{B}_{d,phys}\to TX)+f_s
\;B(\overline{B}_{s,phys}
\to TX) + \nonumber \\
& + & f_{\Lambda_b} \; B(\Lambda_b \to TX) +
\sum \;f_I \;B(I\to
TX) \;\; . \nonumber \\
& & \;\;\;\;\;\;\;\;\;\;\;\;\;\;\;\;\;\;\;\;\;\;\;\;\;\;\;\;\;\;\;
I=\Xi^0_b , \Xi^-_b ,\Omega_b ,B^-_c
\end{eqnarray}
Define $B(\overline{H}_{b,phys}\to TX)$ analogously.
The observed number of primary lepton-$T$ correlations is given by
\begin{eqnarray}
N_{\ell^+ T}& = & N_{b\overline b}\;\alpha_\ell \;\alpha_T
\;B_{s\ell}
\bigg\{\left(1-\chi\right) B\left(H_{b,phys}
\to TX\right)
+ \chi \;B\left(\overline{H}_{b,phys}\to TX\right)\bigg\} \;,
\nonumber
\\
N_{\ell^- T} & = & N_{b\overline b} \;\alpha_\ell \;\alpha_T \;
B_{s\ell}
\bigg\{\left(1-\chi\right) B\left(\overline{H}_{b,phys}\to
TX\right) + \chi \;B\left(H_{b,phys}\to TX\right)\bigg\} \;,
\end{eqnarray}
where $\alpha_\ell$ and $\alpha_T$ are experimental
acceptance/efficiency
factors for $\ell$ and $T$,
$N_{b\overline b}$ is the number of $b\overline b$ events, and
$B_{s\ell}$ is the average semileptonic branching ratio for the
mixture
of $b$-hadrons $H_b$.
The $\chi$ parameter is determined from the dilepton sample, where
the two
leptons come from different $b$-hadrons
\cite{schroder},

\begin{equation}
\frac{N_{\ell^+\ell^+} + N_{\ell^-\ell^-}}
{N_{\ell^+\ell^-} + N_{\ell^+\ell^+} +  N_{\ell^-\ell^-}} =
\frac{2\chi(1-\chi )N_u +\left[\left(1-\chi\right)^2 +\chi^2
\right]
N_\ell}
{N_u + N_\ell}
\end{equation}
Here $N_u (N_\ell )$ is the predicted unlike-sign (like-sign)
dilepton rate in the case of no mixing.
Many systematic errors cancel in the ratio
$N_{\ell^- T} /N_{\ell^+ T}$, which together with
$\chi$ determines the important quantity $I_T$,
\begin{equation}
I_T = \frac{\frac{N_{\ell^-T}}{N_{\ell^+ T}} (1-\chi )-\chi}
{1-\chi- \frac{N_{\ell^-T}}{N_{\ell^+ T}} \chi}\;.
\end{equation}

The dilution for tag $T$ is
\begin{eqnarray}
D_T & = & \frac{B(H_{b,phys}\to TX) -B(\overline{H}_{b,phys}\to TX)}
{B(\overline{H}_{b,phys}\to TX) + B(H_{b,phys}\to TX)} = \nonumber \\
& = & \frac{1-I_T}{I_T +1} \;.
\end{eqnarray}
Note that the dilution for tag $T$ can be determined in more general
situations than that indicated by Eq.~(2.23).
For example, in the case of incoherent $b\overline b$ production,
one may wish to use self-tagging via primary interaction hadrons or a
sophisticated jet charge technique and this can be accomplished via
the $\ell^\pm T$ data sample. In such cases however, the result
does not provide information about $I_T$ because $T$ does not
consist of decay products of the other $b$-hadron in the event.

The $\chi$ parameter is related to the semileptonic branching ratios,
$B^d_{s\ell}$ and $B^s_{s\ell}$,
of the $B_d$ and $B_s$ mesons by,  \begin{equation}
\chi =\frac{B^d_{s\ell}}{B_{s\ell}} \;f_d \;Prob
(B_{d,phys}\to\overline{B}_d )+
\frac{B^s_{s\ell}}{B_{s\ell}} \;f_s \;Prob
(B_{s,phys}\to\overline{B}_s)\;.
\end{equation}
It measures the probability
\begin{eqnarray}
Prob (\overline{H}_{b,phys}\to H_b) & = & f_d \;Prob(B_{d,phys}
\to\overline{B}_d)
+\nonumber \\
& + & f_s \;Prob (B_{s,phys}\to \overline{B}_s )=\chi \;,
\end{eqnarray}
for the case of equal semileptonic branching ratios of the $B_d ,
B_s$ and
$\overline{H}_b$,
\begin{equation}
B^d_{s\ell} = B^s_{s\ell} = B_{s\ell} \;.
\end{equation}

This section described in detail how to determine the important
ratios
$L_T$ and $I_T$, and how to extract the dilution for any tag $T$.
Combining measurements from $\Upsilon(4S)$ and high energy
experiments---that is, $N_{\ell^- T} /N_{\ell^+ T}$ and $I_T$,
respectively---singles out information about inclusive production of
$T$ in $\Lambda_b$ and time-evolved $\overline B_s$ decays versus
$\overline \Lambda_b$ and time-evolved $B_s$ decays.
The next section applies this formalism to existing
$\ell^\pm T$ data samples, and thus informs us about good $b$-tags.

\section{Tags}

As discussed in section I, tagging the flavor of the other $b$-hadron
in
the event must work in principle. In this section we consider a wide
variety of possible tags. In addition to single particle
tags, there exist more general tagging techniques such as
jet charge $Q_j$ which has been shown to be a powerful
tagging tool in the LEP experiments \cite{jetcharge}.
To fully optimize jet-charge tagging however, one should
weight displaced tracks
differently than the tracks associated to the primary interaction
since the fragmentation hadrons are expected to be anti-correlated in
charge
(the basic assumption of self-tagging), while the charge of the final
hadrons
is correlated to the original $B$ flavor. A sophisticated jet charge
algorithm based upon this and other considerations is being
developed for use in
a hadron accelerator environment and is expected to become a very
powerful
tagging technique~\cite{dikss}.  It improves tagging at $e^+e^-$
colliders as well.  Jet charge tagging uses not only
the other $b$-jet in the event, but also the signal $b$-jet when
possible.

The formalism developed in the last section allows one to calculate
$D_T$ from existing $\ell^\pm T$ correlations.
One can propose other good tags, which can be tested with currently
available data sets. Currently all data come from $\Upsilon (4S)$
experiments,
but we expect this to change in the near future.
Tags need not be restricted to specific particle types but may
also be defined by correlating characteristic topologies
of one $b$-jet with the flavor of the other
$b$-hadron in the event. The flavor could be determined for instance,
from the primary lepton, from an inclusive $B^\pm$ data sample, from
an inclusive flavor-specific $\stackrel{(-)}{B_d}$ data sample, from
fully
reconstructed $b$-hadrons, or from topologically separated
$B\overline B$ events at an asymmetric $\Upsilon (4S)$ machine.
Tags at the $\Upsilon (4S)$ are discussed first. High energy
experiments should study the effectiveness of all tags
mentioned for the $\Upsilon (4S)$, but can also expect
additional tags to become available as a result of the
incoherence of the $b\overline b$ pair, as discussed below.

\subsection{List of tags}

Known $\ell^\pm T$ correlations at the $\Upsilon (4S)$ are
summarized in Table I \cite{baryonglasgow}, \cite{crawford}
-\cite{tp}.
Columns I-VI list the tag particle or
particles $T$, literature references, observed numbers of $\ell^\pm
T$
correlations (or a proportional quantity), the probability
of a wrong lepton-charge assignment $p/2$, and the calculated ratio
$L_T$ from Eqs.~(2.16)-(2.17).
If the $\ell^\pm T$ correlation already takes into account the
effects of secondaries $(b \to c \to \ell^+)$, then only $B^0
-\overline{B^0}$ mixing effects need
to be considered, and $p/2 =0.079 \pm 0.009$. On the other hand,
if secondaries have not been dealt with, then we use the ARGUS
estimate of $p/2=0.14\pm 0.02$ for ARGUS data where
$p_\ell > 1.5$ GeV/c \cite{z92}.
Effects of secondaries are much smaller for CLEO
\cite{zoellerprivate},
\begin{equation}
(p/2)_{secondary} = \left\{\begin{array}{lr}
0.028 \pm 0.010 & \;\text{for}\;\; P_\ell >1.4\; GeV/c \nonumber \\
0.020 \pm  0.008 & \;\text{for} \;\; P_\ell > 1.5 \;
GeV/c\end{array}\right\}
\end{equation}
We use those numbers for CLEO results for which all backgrounds were
subtracted, except for secondaries and
$B_d -\overline{B}_d$ mixing.
We do not understand the large discrepancy between ARGUS and CLEO
regarding the effects of secondaries and leave it to be sorted out
among the two collaborations.
We ignore $B_d -\overline{B}_d$ mixing effects upon secondaries,
because they are much smaller than the error on secondaries.
For future $\ell^\pm T$ correlations, one may wish to cut
at a higher lepton momentum where secondaries are negligible.

It has been known for many years that primary leptons
\cite{leptontag},
charged kaons
\cite{ktag}
and charmed hadrons from $b$-decays are good $b$-tags. Table I
reviews the current data on charged kaons and shows that
$K^{*-}, \overline{K}^{*0}, D^0 , D^{*+}, \Lambda~\cite{lambdatag},
\Lambda_c,$ and
$\Lambda\overline p$ identify $b$-flavor well.

Whereas the ratio $L_T$ determines the cleanliness of tag $T$,
the inclusive yield $R_T$ tells us how copious it is. Table II lists
both
quantities and
shows that $K^-$ is to date the most abundant tag, with an
inclusive yield of 0.85 \cite{browder}. The inclusive yields of the
other tags
are 0.15, 0.18, 0.57, 0.24, 0.04, 0.023, 0.06, 0.06
for $\overline{K}^{*0}, K^{*-}, D^0 , D^{*+}, \Lambda , \Lambda
\overline
p ,
\Lambda_c$ and $\overline p$ (not from $\overline{\Lambda}$),
respectively.
Although $\overline p$ is not a good tag {\it per se}, it becomes one
once
$\overline p$ from $\overline{\Lambda}$ are subtracted. This may
be welcome news for CP studies at $e^+e^-$ colliders operating
at, or slightly above, the $\Upsilon (4S)$, since the same charge of
a
primary lepton, $K$ or $p$ (not from $\Lambda$) tags the $B$,
and one could allow for misidentifications among them.

Table III lists a few more decay daughters of $B$ mesons that are
expected to be good flavor tags $(L_T \ll 1)$. Correlating them
with hard, primary leptons needs yet to be performed. Some of them
are copiously produced in $B$ decays. The respective yields of
$D^+ ,D^-_s ,\Xi^0_c, \Xi^+_c$ and $\Xi^-$ are 0.25, 0.12, 0.02,
0.02, 0.003 \cite{browder}.
Due to large uncertainties in the absolute branching fractions
of the modes in which the $\Lambda_c , D^-_s ,\Xi^0_c$ and $\Xi^+_c$
are seen, their inclusive yields in $B$ decays could differ
sizably from the values listed in Tables II and III.
Theoretical considerations lead us to suspect that the $\Lambda_c$
and
$D^-_s$ yields in $B$ decays may well be significantly
underestimated in Tables II - III.  Elevated yields of charmed
hadrons in $B$ decays would resolve the so-called semileptonic
branching fraction puzzle of the $B$ mesons~\cite{bigibaffle}.

A sizable fraction of charged hyperons may live long enough to be
detected via $dE/dx$ without full reconstruction~\cite{ikss}.
Whereas
the $\Xi^-$ and $\Omega^-$ are probably good tags, the situation
pertaining to $\Sigma$'s is less clear \cite{sigma}.  Because $dE/dx$
is not able to discriminate among the various charged hyperon
species, experimental studies will be necessary to determine the
effectiveness
of this tag.

We predict that most $D^-_s$ come from the virtual
$W$ with charge opposite to that of most charged $D$'s in $\overline
B$
decays. A simple tag based, for example, upon the charge of the
tertiary
vertex will therefore not succeed because of similar inclusive
production
rates of $D^-_s$ and $D^+$ in $\overline B$ decays.
One could however discriminate inclusively between these charmed
hadrons in at least three ways. First, the $D_s$ lifetime is 2.3
times shorter than that of the $D^+$. Second, the momentum spectra of
the
two charmed mesons differ \cite{browder}.
In $B$ decays, the $D^{(*)}_s$ is generically produced in association
with
another charmed meson and is seen in two body modes about $50$
percent of the time~\cite{menary}, in contrast to the $D$ mesons.
Whereas the spectrum of the $D_s$ is peaked at high momenta,
that of the $D$ meson is much flatter~\cite{browder,menary}.
Third, $K/\pi$ separation discriminates
between the two charmed mesons
because the $D^+_s$ is mainly seen in $S=0$ final states
containing an even number of kaons whereas the $D^+$ decays mainly in
$S=- 1$
final states containing an odd number of kaons.
One has to take into account, however, that the Cabibbo-suppressed
modes are anomalously large for the $D^+$.

Further, consider a displaced vertex with a few charged tracks which
is consistent
with being a charmed meson (or even a $\Lambda^+_c$). If
two of the tracks satisfy a $\phi$ hypothesis, then it is
probable that the parent is a $D_s$, with charge
determined from the other track(s). The inclusive yield of
$\phi$ in $D_s$ decays is quite enhanced over that in $D^+$ and $D^0$
decays \cite{exclusivesum}.  If no two tracks satisfy the $\phi$
hypothesis, one could search for a $\overline{K}^{*0}$ analogously.
The inclusive yield of $\overline{K}^{*0}$ in $D$ meson decays
dominates that of $D_s^+$ decays \cite{exclusivesum}.  Thus,
$\overline{K}^{*0}$ would tag $b$-flavor well. A systematic study of
all such correlations is currently underway and will likely enlarge
future data samples \cite{workinggroup}. Clearly, it will be
useful for experiments to measure the inclusive
yields of $\phi , \overline{K}^{*0} ,{K}^{*0},{K}^{*-},{K}^{*+}$
in $D^+_s , D^+, D^0,$ and $\Lambda_c$ decays.

Pions are by far the most copious type of charged particles in $B$
decays
with an average multiplicity of about 4 per $B$ decay \cite{browder}.
Thus, any characteristic of a charged $\pi$ (momentum, $P_{T,rel}$,
etc.)
from one $b$ which can be found to
exhibit a strong correlation with the charge of the hard, primary
lepton from the semileptonic decay of a partner $b$ could also be
employed as a tag. Such characteristics
could also be searched for in the data sample where one $B$ has been
either fully reconstructed or spatially disentangled.

At least in the case of the $\overline p$, it is possible to turn a
bad tag
with $L_T\approx 1$, into a good one.
For instance, a $T=\overline p$ may become a better tag
when associated
with a $K^+$ or $\pi^+$ from the same $T$ vertex. Recall that a
$\overline p$ becomes a great tag when associated with a $\Lambda$.
One could try and make particle associations for other marginal tags
or
one could measure $L_T$ as a function of
momentum to see if there exists a momentum range in which $T$ tags
the
$b$-flavor much better.
At an asymmetric $\Upsilon (4S)$ factory and at high energy
experiments
the purity of single particle
tags is generally enhanced by demanding that they originate from
displaced
vertices. We would like to reiterate however that not only single
particle
tags should be used, but any conceivable event
topology should be studied for strong correlations with $b$-flavor.
We are confident that such a program will find many additional tags.

While most good tags at the $\Upsilon (4S)$ remain good ones (and
some even become better ones) at
high energy experiments and should be vigorously investigated,
there are exceptions, such as $D^-_s$ and $\overline p$ (not from
$\overline{\Lambda}$). As mentioned above,
additional selection criteria may be able to turn even the
exceptions into usable tags. At the $\Upsilon (4S)$,  only the
$B^\pm$ and $\stackrel{(-)}{B_d}$ species are created and
$D^-_s$ for these cases originates mainly from virtual $W\to
\overline c s$
decays, with an inclusive yield of 12\%. In contrast, the $b\to c$
transition---responsible for almost all $b$-decays---governs the
inclusive $D^+_s$ production in $\overline{B}_s$ decays at high
energy experiments. There is an
additional contribution of about ten percent from oppositely
charged $D^-_s$ originating from virtual $W\to\overline c s$ decays.
Because of the expected production fraction of $B_s$ mesons, the
yields
of $D_s$'s from the virtual $W\to \overline c s$ and from the $b\to
c$ transition are comparable and the large $B_s -\overline{B_s}$
mixing
washes out the initial flavor information.  Consequently,
the $D^{\pm}_s$ tags are not as clean as at the $\Upsilon (4S)$,
because of the
$\stackrel{(-)}{B_s} \to D_s^\pm X$ background.
However, separating $D_s$ originating from $W\to \overline c s$
versus
$b\to c$ decays may still be possible due to their different momentum
spectra.
The $D^+_s$ momentum spectrum for $D^+_s$ coming from the $b\to c$
transition is expected to be similar to that of the $D$ meson in
$\overline{B}$ decays---that is, much flatter than the high
momentum-peaked spectrum of $D_s$ originating from the virtual $W$
\cite{browder}.

A potentially more severe problem exists for the prompt proton
(i.e. not from $\Lambda$),  in $b$-decays.
If one were able to separate the yield of ``prompt protons"
into those
from $B$-mesons and those from
$\Lambda_b$'s, then the two yields could be used as good tags.
Indiscriminate use of $p$ (not from $\Lambda$) will not
tag $b$-flavor well. We know of no ingenious method to accomplish
this and can offer only a couple of, at best, marginal suggestions.
One is to seek an additional antiproton. Distinguishig $p\overline p
X$
events from $pX$ events may allow one
to enrich the $p$ from $\Lambda_b$ data sample. Alternatively,
perhaps the
momentum spectrum of the prompt protons discriminates between the two
$b$-sources. Lastly, one could try to exploit the slight difference
in the lifetime of $\Lambda_b$'s $(\sim 1.2
ps$) versus $B$-mesons $(\sim 1.6 ps$) \cite{sharma}.
Regardless of whether or not detached protons are good
tags, they could be used as a potent $b$-trigger~\cite{workinggroup}.
Their inclusive yield in $b$-decays is substantial, and the
background from charmed baryons can be disentangled due to their much
shorter lifetimes.

High energy experiments could identify the $b$-flavor using a
sophisticated jet-charge
$Q_j$ \cite{dikss}. They could unambiguously determine the
flavor of the
accompanying $b$-hadron, either from its charge $B^\pm$ or from it
being a $b$-baryon versus $\overline b$-baryon
\cite{dr}.
Self-tagging may be able to distinguish a $B$ from a $\overline B$ by
correlating the beauty meson with the charge of a primary hadron
nearby in phase space \cite{gnr,gr}.
If self-tagging were to work for neutral $B$'s it could be applied
directly to signal $B$ hadrons.
Even if it does not work for neutral $B$'s, it may work for
charged $B$'s, in which case the self-tagging scheme could be turned
``literally" upside-down to tag a signal $B$ hadron
via the charge
of a primary hadron found in a small cone about the axis of
a jet opposite to it.
This works well for $Z^0 \to b\overline b$
where the two $b$-jets are generically back
to back, but probably needs to be augmented for  $p\overline p \to
b\overline b +\ldots
\;,\;$  by requiring a displaced vertex to define the opposite
jet~\cite{dikss}.
This primary hadron tag could then be combined with other
information to augment the tag. For instance, identifying the charge
of the other
$B^\pm$ could be enhanced by correlating it with the opposite charge
of a primary hadron nearby its phase space.  Experiments will
determine the optimal
tags which combine information
from the other  $b$-jet with a primary hadron nearby its phase space.
This is in fact an example of a more general
and sophisticated jet charge tag.

There are many promising event topologies for tagging which can be
identified
by a systematic analysis of inclusive $b$-decays.  Since
almost all $b$-decays involve the $b\to c$ transition, their detailed
understanding requires extensive knowledge of charmed hadron decays.
We
have therefore been studying all aspects of charm decays, such as
inclusive decays, exclusive decays, and theoretical constructs and
these
will be reported elsewhere, when a thorough analysis has
been completed \cite{workinggroup}.  Here we restrict ourselves to a
few examples. For a $2d \; (3d)$ vertex detector, one possible
event
topology could be a detached vertex with $3 \; (2)$ or more charged
tracks.   Suppose that neither leptons nor protons are seen and that
the vertex is consistent with having been  formed by a charmed hadron
decay.
The number
of kaons in the event then discriminates between $D^+, D^0$ and
$D_s^+$.
Good particle identification would make this information accessible
to an
experimentalist. In the absence of clean particle identification, one
could
weight tracks by the probability that they correspond to a given
particle type.
Such a vertex could also arise from decays involving a lepton. For
example,
$$\overline{B}\to D^{*+} \;\left[\to\pi^+ D^0 \left(\to K^-
\pi^+ X\right)\right]\ell^- X \;\;\text{or}$$
$$\overline{B}\to D^0 (\to K^- \pi^+ X) \ell^- X,$$
where the short $D^0$ lifetime would likely result in a merging
of the tertiary $D^0$
vertex with the secondary $b$ decay vertex. (Future high resolution
experiments could separate the tertiary
from the secondary vertices to enhance tagging.) Of course, the
charge of
the lepton as well as that of the kaon correlate well with the
$b$-flavor.

Another event topology could be a displaced lepton which does not
associate with a  $3 \; (2)$  or
more charged track vertex consistent with a charmed hadron
\cite{ikss}.  The charge of the lepton would be a good $b$-tag, and
the other vertex could reveal information upon the nature
of the charmed hadron that could be used to corroborate the lepton
tag.
The vertex mass technique discussed in Section I could be turned into
a tag by utilizing the probabilities of particle identifications, the
charge of the partially reconstructed $b$-hadron, and any other
available, discriminating information in the event.
Further, it has been suggested that it may be possible to accumulate
large inclusive
$b$ hadron data samples~\cite{ikss} without requiring leptons in the
final
state by selecting dijet events with displaced vertices in both jets
together with some minimal requirement to reject charm (such as
vertex mass discussed above). Such samples could be an interesting
source
for a variety of studies of non-leptonic $b$ decays.

Although quite a few tagging schemes have been discussed, we are
confident
that an intense study would find many more usable tags
for any given detector.
Clearly, to optimize tagging, we advocate a judicious dilution
weighted
combination (which avoids multiple counting) of all usable
tags.  This implies a well understood dilution for each tag $T$,
which fortunately can always
be determined from either $\ell^\pm T$ (last section)
or $TB^\pm$ correlations to which we now turn.

\subsection{$T B^\pm$ correlations}

Consider incoherent
production of $b\overline b$ events at a high energy experiment.
Define
the
number of $B^\pm$ events correlated with a given tag $T$ or
$\overline T$
by
\cite{dilution}
\begin{eqnarray}
P_1 & \equiv & N(TB^+) \;, \;\;\;\;P_2 \equiv N(TB^- ) \;. \nonumber
\\
P_3 & \equiv & N(\overline T B^- ) \;, \;\;\;\;P_4 \equiv N(\overline
T
B^+ )\;.
\end{eqnarray}
Statistics can be doubled, because for charge symmetric production of
$b\overline b$ events---as in $p\overline p \to b\overline b +\cdots$
or
$e^+e^- \to Z^0 \to b\overline b$---we get
\begin{equation}
P_1 = P_3 \;, \;\;\;P_2 = P_4 \;.
\end{equation}
As for inclusive $B^\pm$ data samples, a few suggestions were
mentioned
in Section I, which we repeat here.

One could use the displaced $J/\psi$ data sample paired with an odd
number
of charged prongs originating from the same $J/\psi$ vertex.
The detached $J/\psi$ guarantees $b$-parentage. No particle
identification
is required, and missing neutrals pose no problem. This is a clear
$B^\pm$
data sample.
Alternatively one could use $D^{(*)} \ell^- X$ events that are
consistent
with coming from a $b$-decay or the fully hadronic sample mentioned
above.
(For the latter two however the backgrounds due to collinear
$c\overline c$ production at hadron machines must be taken into
account.)
If
self-tagging works for charged $B$'s,
one may wish to pair the charged $B^\pm$ data sample with oppositely
charged hadrons nearby in phase space to reduce backgrounds.
The dilution of tag $T$ is
\begin{equation}
D_T = \frac{P_1 -P_2}{P_1 +P_2}
\end{equation}
and could be checked against the result obtained from
$\ell^\pm T$ correlations (see Section II).
(Of course, when the tag $T$ consists of decay products of the other
(non signal) $b$-hadron, the $T B^\pm$ correlations determine again
$I_T$.)
Further cross-checks
for $D_T$ involve the smaller samples of fully reconstructed
$B^\pm$, such as $B^\pm \to J/\psi K^\pm ,J/\psi K^{*\pm},$ and
flavor-specific $\stackrel{\textstyle{(-)}}{B_d}$.
Assuming incoherence, the same dilution occurs for the neutral
$B^0$ mesons ($B_d$ or $B_s$), once $B^0 -\overline{B^0}$
mixing has been removed, i.e.,
\begin{eqnarray}
N(TB^0) \sim ~P_1 \;,& & \; N(T\overline{B^0}) \sim~ P_2 \nonumber \\
 N(\overline{T}\;\overline{B^0}) \sim ~P_3\;, & & \;
N(\overline{T} B^0 ) \sim ~P_4.
\end{eqnarray}
Here $B^0$ and $\overline{B^0}$ indicate the initial flavor of the
neutral $B$ prior to mixing.
This fact allows one to compare $D_T$
measurements from a variety of data samples involving $B_d$
mesons. It also allows one to measure $B_s -\overline{B}_s$ mixing
and to study CP violation, which will be the topic of the next
section.

Consider flavor-specific modes of neutral $B$ mesons, such as
$$
B_d \to J/\psi K^{*0} , \; \overline{D}^{(*)} \ell^+ X, D^{(*)-}
\pi^+, \overline{D}^{(*)}X_{u \overline d},\;\;
$$
$$
\overline{B}_d \to J/\psi \overline{K}^{*0}, \; D^{(*)} \ell^- X \;,
\; D^{(*)+} \pi^- \;, \; D^{(*)} \overline{X}_{u \overline d},
$$

$$
B_s \to D^-_s \ell^+ X, J/\psi\overline{K}^{*0}, D^-_s \pi^+, D^-_s
X_{u \overline d},
$$
$$
\overline{B}_s \to D^+_s \ell^- X, \;\;J/\psi K^{*0} \;, D^+_s \pi^-,
D^+_s \overline{X}_{u \overline d},
$$
where the flavor of $K^{*0}(\overline{K}^{*0})$ is identified by
the charge
of their charged daughter-kaon. The symbol $X_{u \overline d}$
represents
a collection of particles with zero strangeness, such that the only
consistent underlying quark-transition for $\overline{D}^{(*)}X_{u
\overline d}$ and/or $D^-_s X_{u \overline d}$ final states is
$\overline b \to \overline c u \overline d$ and not $\overline b \to
\overline c c \overline s$.  Define a ``right-sign"
combination as either
$T B^0$ or $\overline T \;\overline{B}^0$, and a
``wrong-sign"
combination by $T\overline{B}^0$ or $\overline T B^0$. The
time-dependence
of the relative numbers of the right-sign $R$ and wrong-sign $W$
combinations
are
\cite{dilution}:

\begin{eqnarray}
R(t) & = & e^{-\Gamma t} \bigg\{ P_1 \cos^2 \frac{\Delta mt}{2} +P_2
\sin^2
\frac{\Delta mt}{2}\bigg\} \;,\nonumber \\
W(t) & = & e^{-\Gamma t} \bigg\{ P_1 \sin^2 \frac{\Delta mt}
{2} +P_2 \cos^2 \frac{\Delta mt}{2}\bigg\} \;.
\end{eqnarray}
The time-dependent asymmetry is then

\begin{equation}
\frac{R(t) -W(t)}{R(t) +W(t)} =D_T \cos \Delta mt \;,
\end{equation}
which integrates to

\begin{equation}
\frac{\int dt \;[R(t) -W(t)]}{\int dt \;[R(t) +W(t)]}= D_T \;
\frac{1}{1+x^2} \;,
\end{equation}
with

\begin{equation}
x\equiv (\Delta m/\Gamma )_{B^0} \;.
\end{equation}
(Note that the above equations assume equal lifetimes for the heavy
and light
mass eigenstates of $B^0$ which is an excellent approximation
for the
$B_d$-system, but may be violated at the 10-20\% level for the
$B_s$-system
\cite{Bslifedifference}.)

Since $B_{d}-
\overline{B}_d$ mixing
is known
\cite{schroder},
$x_d =0.71\pm 0.07$, Eqs. (3.7) - (3.8) imply that dilutions $D_T$
can also be measured with flavor-specific
$B_d$-
modes (both time-dependent and time-integrated).
Furthermore $B_{s}-
\overline{B}_s$
mixing could, for instance, be measured via the time-evolution of
flavor
specific modes of $B_s$ correlated with tags $T$.

Finally, suppose that in general one were to identify a particularly
clean and copious tag $T'$, where the
dilution
$D_{T'}$ is known from either $\ell ^\pm T'$ or $B^\pm T'$ or
flavor-specific $\stackrel{(-)}{B_d} T'$
correlations or
any combination thereof. Then, $D_T$ for some other tag $T$ could be
determined from
$T\overline{ T}'$
and $TT'$ correlations. The most
accurate determination of $D_T$ is obtained by correctly
weighting and combining all known $T \stackrel{(-)}{T'}, TB^\pm$,
flavor specific $\stackrel{(-)}{B_d} T$ and
$\ell^\pm T$ correlations.

We know that tagging the flavor of the other (non signal) $b$-hadron
in the
event must work in principle. We discussed many possible
tags and recommended that one seek and find event topologies
that can be used as tags. The dilution for any tag $D_T$ can be
determined from $B^\pm T$ correlations. It is also obtained
from primary lepton-tag correlations and
$\stackrel{\textstyle{(-)}}{B_d} T$
data samples, which could serve as cross-checks and be combined for a
more
accurate final determination.
The correct dilution weighted combination of all possible tags (which
avoids
multiple counting) optimizes tagging. It allows studies of
$B_s -\overline{B}_s$ mixing by measuring for instance the
time-evolution of
flavor-specific $B_s$ modes. It also allows
CP studies to be contemplated which is the topic of
Section IV.

\section{CP Violation}

Of central importance is the measurement of CP violation and the
clean
extraction of the weak phases, to which we now turn. We denoted by
$B^{0}
_{phys}$ a time-evolved state which was initially pure $B^0$,

\begin{equation}
|B^{0}_{phys}(t=0)\rangle=|B^0\rangle\;.
\end{equation}
 $\overline {B}^{0}_{phys}$ was defined
analogously. Consider the case where the neutral $B$ is seen in a
CP-eigenstate $f$,
such as $B_d\rightarrow J/\psi K_S,\pi^+\pi^-, B_s \rightarrow
D^{+}_{s}D^{-}_{s},
J/\psi\phi$ \cite{psiphi}.
The time-dependent or time-integrated CP-violation asymmetry is
given by

\begin{equation}
A_f \equiv \frac{\Gamma (B^0_{phys} \to f) -\Gamma
(\overline{B}^0_{phys}
\to f)}{\Gamma (B^0_{phys} \to f) +\Gamma (\overline{B}^0_{phys}\to
f)}\;,
\end{equation}
where the time-dependent widths are

\begin{equation}
\stackrel{\textstyle{\!\!\!(-)}}{\Gamma (t)} \equiv \Gamma (
\stackrel{\textstyle{\!\!\!\!\!\!\!(-)}}{B^0_{phys}}
(t) \to f) \sim e^{-\Gamma t} \bigg\{ 1
\stackrel{\textstyle{(+)}}{-}
Im \lambda \;\sin \;\Delta mt\bigg\} \;.
\end{equation}
Table IV lists the interference terms $Im\lambda$ in terms of the
angles of
the CKM
unitarity triangle \cite{unitarity}. The $J/\psi K_S$ asymmetry
determines $\sin 2\beta$,
and $\pi^+\pi^-$ determines $\sin 2\alpha$ if penguin  diagrams
can be neglected.
The $B_s \to J/\psi \phi$ asymmetry measures the angle
$\gamma$
once $\mid V_{ub} / V_{cb}\mid$ is known
\cite{snowmass}.
The time-integrated asymmetry is

\begin{equation}
A_f = \frac{\int dt \left[\Gamma \left(t\right)
-\overline{\Gamma} \left(t\right)\right]}
{\int dt \left[\Gamma \left(t\right) +\overline{\Gamma}
\left(t\right) \right]}
=\frac{-x}{1+x^2} Im\lambda \;.
\end{equation}

Time-dependence is not crucial for the $B_d$ meson $(x \approx 0.7)$,
while
it is crucial for the $B_s$ meson, where $x\gg1$ is expected
\cite{schroder,al} and observed \cite{sharma}.
Several experiments will be able to study time-evolution, which
therefore should
be done. The time-dependent asymmetry is

\begin{equation}
A_f (t) \; = \;\frac{\Gamma(t) -\overline{\Gamma}(t)}
{\Gamma (t) + \overline{\Gamma} (t)}
 =-Im\lambda \;\sin (\Delta mt) \;.
\end{equation}

Our imperfect knowledge of the initial $b$-flavor introduces
dilution $D_T$.
Correlating the neutral $B$ mode $f$ with tag $T$, we obtain the
observed asymmetry,

\begin{equation}
A^T_f \equiv \frac{N({T} ,f) - N(\overline T, f)}
{N(T ,f) + N(\overline T,f)}\;.
\end{equation}
It is related to the true asymmetry $A_f$ by

\begin{equation}
A^T_f = D_T \;\;A_f \;,
\end{equation}
which holds for both time-dependent and time-integrated studies.
Whereas the true asymmetry $A_f$ is independent of the tag $T$, the
observed asymmetry $A^T_f$ and dilution $D_T$ depend on $T$.
The correct dilution weighted combination of all accessible tags $T$
optimizes the measurement of the true asymmetry $A_f$, and
hence of
the interference term $Im \lambda$ which determines the relevant weak
phase.

Many exclusive modes measure the same unitarity angle within the CKM
(Cabibbo-Kobayashi-Maskawa) model of CP violation \cite{ckm}.  They
can be added to increase statistics.  The addition must be done
carefully lest a partial cancellation of the asymmetry due to CP-even
and CP-odd modes \cite{bigi} makes the extraction of the relevant
unitarity angle less crisp.  For instance, the angle $\beta$ can be
determined also from
\begin{equation}
B_d \to J/\psi K^{*0} (\to \pi^0 K_S), J/\psi \rho^{0}, J/\psi
\omega, \cite{kkps,dqstl}
\end{equation}
\begin{equation}
B_d \to D \overline{D}, D^* \overline{D}, D \overline{D}^*,  D^*
\overline{D}^*, \cite{rosnerDD,adkl,aleksanDDbar}
\end{equation}
\begin{equation}
\text{and}\;\;\;\;\;\;  B_d \to \overline{D}^0 (\to f_{CP}) X_{u
\overline d} \;, \cite{dsnyder}.
\end{equation}
Here $X_{u \overline d}$ denotes a collection of particles which
guarantee that the $B_d \to \overline{D}^0 (\to f_{CP}) X_{u
\overline d}$ process is by far dominated by the underlying
$\overline b \to \overline c u \overline d$ quark transition.  The
symbol
$f_{CP}$ stands for $D^0$  decay modes which are either CP
eigenstates
or which can be decomposed into CP eigenstates.  For example, the
decomposition can be accomplished through an angular correlation
study \cite{dqstl}, such as for $D^0 \to \rho^0 \overline{K}^{*0}
(\to K_S \pi^0)$.  The summation of all of these exclusive modes may
well provide the necessary increase in statistics to rule out or
observe CP violation in the CKM model.
Once sufficient statistics have been accumulated one may wish to
undertake precision studies of the CKM model by studying the
CP-violating asymmetries for each of the underlying
quark-subprocesses separately.

Some CP-noneigenstates, such as $B_s \to
D^\pm_s K^\mp$
\cite{adk},
$ \stackrel{\textstyle{(-)}}{D^0} \phi$
\cite{gldphi},
are expected to show large
time-dependent
CP-violation effects and allow a clean extraction of the CKM
unitarity
angle $\gamma$. The
formalism developed for CP-eigenstates can be trivially extended to
include
the case of non-CP eigenstates \cite{adk,adkl,gr}. The algebra
though will
be more cumbersome.

\section{Summary}
Distinguishing $B$ from $\overline B$ is crucial to a deeper
understanding of
nature. Studies of CP violation, $B_s-\overline B_s$ mixing, and
measuring
the
production fraction of flavor tags $T$ from $b$-hadrons versus
$\overline
b$-hadrons
becomes
possible when $B$ and $\overline B$ can be distinguished. In
principle,
tagging the flavor of the
other $b$ in the event achieves this goal.

An \lq\lq ideal detector'' could use almost any $b$-decay as a flavor
tag
with an overall  dilution of

\begin{equation}
D\equiv \frac{\text{Prob} (H_{b,phys}\to H_b ) - \text{Prob}
(H_{b,phys}
\to \overline{H}_b )}
{\text{Prob} (H_{b,phys}\to H_b ) +\text{Prob} (H_{b,phys}\to
\overline{H}_b)}  \approx 0.74 \;.
\end{equation}

The existing $\Upsilon(4S)$ data on charged, primary $\ell^\pm T$
correlations---where one $B$ gives rise to the lepton and the other
to $T$---identifies
many good tags.
We have discussed in detail the correct removal of $B_d-\overline
B_d$ mixing effects.

After removing $B_d-\overline B_d$ mixing, we calculated the
important ratio $L_T$.
Because $L_T$ and $I_T$ are crucial for a deeper understanding of
$b$-decay
mechanisms,
we recommend $\ell^\pm T$
correlation be measured
whenever possible, both at $\Upsilon(4S)$ factories and at high
energy machines. We determined the ratio $I_T$
from $\ell^\pm T$ correlations at higher energy machines, where
$b\overline b$
production is incoherent. Because the probability of a time-evolved
$H_b$ to
be seen as its antiparticle is small
$Prob(H_{b,phys}\to \overline{H}_b)\approx 0.13$, the ratio $I_T$
tells one about the relative strength of the inclusive yield of tags
$T$
from
$\overline b$-hadrons versus $b$-hadrons. The dilution $D_T$ of each
tag can
be
determined from $\ell^\pm T$ correlations, and also from $TB^\pm$
correlations. The $B^\pm$ data sample could be simply $J/\psi
X^\pm$ events,
where the
$J/\psi$ is displaced and $X^\pm$ is a collection of charged
particles
originating from the $J/\psi$ vertex.

One should use all conceivable tags, properly weighted and combined,
and not restrict oneself to traditional
primary leptons and
$K^\pm$'s.  An optimal tagging scheme uses all available
information for a given $b$ decay to
weight and sum charges and particle identification
probabilities (and any other pertinent information),  using
different weights for displaced particles than for primary hadrons
\cite{ikss}.  Section III lists many tagging possibilities, and many
more will be presented once a systematic analysis has been
completed \cite{workinggroup}.

Clearly,
one must develop an intimate knowledge of one's detector and
understand its capabilities and limitations, in order to
determine all possible tags and their dilutions $D_T$. Once this is
accomplished, the study of CP-violation, of $B_s- \overline{B}_s$
mixing and the determination of $I_T$ could be
simple exercises in combining correctly all possible
tags correlated with the relevant signal. It is a
challenge worth accepting.

After completion of this report, we learnt about Ref.~\cite{enomoto}
which is of interest to the reader and partially overlaps with an
independent analysis~\cite{dikss}.

\section{Acknowledgements}

We are delighted to express our gratitude to Joe Incandela, Eric
Kajfasz,
Rick Snider and Dave Stuart for very
informative discussions.  We thank Ed Thorndike
for informing us about the existing literature
of lepton-kaon and lepton-$D$ correlations, and James D. Bjorken, Tom
Browder and
Peter S.~Cooper for discussions.  We value Jon Rosner's comments
given on two earlier drafts, and Angie Greviskes for typing the
earliest one.  We are very grateful to Lois Deringer for having done
a marvellous job in turning our scribbles into legible form, and
Sonya Wright for typing the second version.     This work was
supported by the Texas National Research
Laboratory
Commission (the Texas agency for the Superconducting Super Collider),
Fellowship No.~FCFY9303,
and the Department of Energy, Contract No.~DE-AC02-76CHO3000.

\begin{table}

\caption{Primary lepton-tag correlations.  Columns I-VI list the
tag $T$, reference, observed number of $\ell^\pm T$ correlations
(or a proportional quantity), the probability of wrong lepton-charge
assignment $p/2$,
and the calculated ratio $L_T$.}

\begin{tabular}{|l|l|c|c|c|c|}
$T$ & References & $N_{\ell^+ T}$ &  $N_{\ell^- T}$ & $p/2$ & $L_T
\equiv
\frac{\Gamma (B\to TX)}{\Gamma (\overline{B}\to TX)}$ \\
\hline
$\Lambda$ & CLEO\cite{crawford} & 103.0 $\pm$ 12.1 & 31.4 $\pm$ 8.2
& 0.11 $\pm$ 0.01 & 0.19 $\pm$ 0.09 \\
$\Lambda$ & ARGUS\cite{z92} & 55$\pm$13 & 30$\pm$10
& 0.14 $\pm$0.02 & 0.42 $\pm$ 0.26 \\
$K^-$ & CLEO\cite{tipton,thorndike}
& 0.66 $\pm$ 0.05 $\pm$0.07 & 0.19 $\pm$0.05$\pm$0.02
& 0.10 $\pm$0.01 & 0.18 $\pm$0.09 \\
$K^-$ & ARGUS\cite{cronstrom,desy93} & 0.620 $\pm$0.013$\pm$0.038
& 0.165$\pm$0.011$\pm$0.036 & 0.079$\pm$0.009 & 0.18$\pm$0.07 \\
$\overline{K}^{*0}$ & ARGUS\cite{cronstrom} &
0.143$\pm$0.019$\pm$0.012
& 0.014$\pm$0.021$\pm$0.011 & 0.079$\pm$0.009 & 0.01$\pm$0.17 \\
$K^{*-}$ & ARGUS\cite{cronstrom} & 0.169$\pm$0.056$\pm$0.036
& 0.015$\pm$0.049$\pm$0.027 & 0.079$\pm$0.009 & 0.00$\pm$0.34 \\
$D^{*+}$ & ARGUS\cite{schafer} & 28.2$\pm$6.1$\pm$0.9 &
5.5$\pm$4.0$\pm$1.2
& 0.079$\pm$0.009 & 0.11$\pm$0.16 \\
$D^0$ & CLEO\cite{tp} & 0.74 $\pm$ 0.20 & 0.18$\pm$0.21
& 0.10$\pm$0.01 & 0.14$\pm$0.30 \\
$\overline p$ & ARGUS\cite{z92} & 453 $\pm$34 & 333$\pm$34
& 0.14$\pm$0.02 & 0.65$\pm$0.12 \\
$\Lambda\overline p$ & ARGUS\cite{z92} & 27$\pm$6
& 4.5$\pm$3.5 & 0.14$\pm$0.02 & 0.00$\pm$0.14 \\
$\Lambda_c$ & CLEO\cite{baryonglasgow} & 139 $\pm$ 16 &
38$\pm$16 & 0.075$\pm$0.016 & 0.20$\pm$0.13$\pm$0.04

\end{tabular}

\end{table}

\begin{table}

\caption{Inclusive yields in $B$ decays [20]
and their
fractional yields from $B$ versus $\overline B$ mesons.}

\begin{tabular}{|l|c|c|}
$T$(tag) & $L_T \equiv \frac{B(B\to TX)}{B(\overline{B}\to TX)}$
& $R_T \equiv B(B\to TX)+B(\overline{B}\to TX)$  \\
\hline
$K^-$ & 0.18$\pm$0.07 & 0.85$\pm$0.07$\pm$0.09 \\
 &   & Multiplicity: 0.78$\pm$0.02$\pm$0.03 \\
$\overline{K}^{*0}$ & 0.01$\pm$0.17 & Multiplicity:
0.146$\pm$0.016$\pm$0.020 \\
$K^{*-}$ & 0.00$\pm$0.34 &  Multiplicity: 0.182$\pm$0.054$\pm$0.024
\\
$\Lambda$ & (CLEO value) 0.19$\pm$0.09 & 0.040$\pm$0.005 \\
$\Lambda \overline p$ & 0.00$\pm$0.14 & 0.023$\pm$0.004$\pm$0.003 \\
$D^{*+}$ & 0.11$\pm$0.16 & 0.237$\pm$0.023$\pm$0.009 \\
$\overline p$ & 0.65$\pm$0.12 & 0.08$\pm$0.005 \\
$D^0$ & 0.14$\pm$0.30 & 0.567$\pm$0.040$\pm$0.023 \\
$\overline p$(not from $\overline{\Lambda}$) & 0.22$\pm$0.12 &
0.056$\pm$0.007 \\
$\Lambda_c$ & 0.20$\pm$0.13$\pm$0.04 & 0.064$\pm$0.013$\pm$0.019

\end{tabular}
\end{table}

\begin{table}

\caption{Expected good tags $T (L_T \ll 1$) and their
inclusive yields in $B$ decays.}

\begin{tabular}{|l|c|c|}
$T$(tag) & Ref.
& $R_T \equiv B(B\to TX)+B(\overline{B}\to TX)$ [in \%] \\
\hline
$D^-_s$ & \cite{menary} & 11.81 $\pm$0.43$\pm$0.94 \\
$D^+$ & \cite{browder} & 24.6$\pm$3.1$\pm$2.5 \\
$\Xi^+_c$ & \cite{baryonglasgow} & 1.5$\pm$0.7 \\
$\Xi^0_c$ & \cite{baryonglasgow} &  2.4$\pm$1.3 \\
$\Xi^-$ & \cite{browder} & 0.27$\pm$0.06 \\
$\Omega^-$ &  &  \\
$\Sigma$ &  &  \\
$\vdots$ &  & \\
Charge of characteristic $\pi$ & \cite{browder}
& Multiplicity: 3.59$\pm$0.03$\pm$0.07 \\
(not from $\Lambda, K_S$) &  &  \\

Characteristic event topology & &

\end{tabular}
\end{table}

\begin{table}

\caption{A few representative modes and their interferences
$Im \lambda$ given in terms of the angles of the unitarity triangle.
Here
$\theta_c \approx 0.22$.}

\begin{tabular}{|c|c|}
Mode & $Im\lambda$ \\
\hline
$B_d \to J/\psi K_S$ & $\sin (2\beta )$ \\
$B_d \rightarrow \pi^+\pi^-$ & $\sin (2\alpha )$ \\
$B_s \to D^+_s D^-_s , J/\psi\phi$\cite{psiphi}
& $2\;\theta_c\; \mid V_{ub} / V_{cb} \mid\sin\gamma$

\end{tabular}
\end{table}


\begin{references}
\bibitem{incoherence}
Incoherence of $b\overline b$ events can be
guaranteed for the method which tags the flavor of the other
$b$-hadron by demanding many primary hadrons between the two
$b$-hadrons as would be expected for well separated jets.
There is a very slight chance that this assumption breaks down
for the self-tagging scheme. If coherence effects were to occur it is
of value to understand them, so that the data sample
of tags can be increased. Some work was done by
M. Gronau and J. L. Rosner, Phys.Rev.Lett. {\bf 72}, 195 (1994),
and additional work is desirable.
\bibitem{b0id}
At an asymmetric $\Upsilon (4S)$, one could use
in addition partially reconstructed $B$ events where the $B \overline
B$ events are topologically separated. Consider the case in which the
net
charge of
displaced prongs from one $B$ is zero. Then, for
instance,
the charge of a primary lepton identifies the $b$-flavor of that
neutral
$B$ meson at the time of its decay.
\bibitem{strictly}
Eqs. (2.3)-(2.4) strictly hold for $B^0$
and $\overline{B^0}$ data samples which consist of modes that can
be fed only from a neutral $B$ but
\underline{never} from its antiparticle---for instance, the
semileptonic $B$-modes, or $D^{(*)-}D^{(*)+}_s$, or $J/\psi K^{*0}
(\to K^+ \pi^-)$.
Subtle interference effects occur for
$B^0$ modes that can be fed from both $B^0$ and $\overline{B^0}$.
Since those effects are well understood \cite{dthesis}, the data
sample increases by including modes which originate mainly from one
neutral $B$ and rarely from its antiparticle.
\bibitem{sigma}
In $\overline B$ meson decays the $\Sigma$ could arise from the
subsequent decay of a charmed baryon which is seen in a $\Sigma$.
However, there is another contribution which leads to
anti-$\Sigma$'s.  The dominant $b \to c \overline u d$ transition
could be accompanied with $s \overline s$ fragmentation, which leads
to $\Xi^{(r)}_c \overline{Y}^{(r)}$ modes.  The superscript $(r)$
denotes the groundstate or any of its resonances.  The
$\overline{Y}^{(r)}$ could be seen as $\overline \Sigma$'s.  Primary
lepton-$\Sigma$ correlations determine the ratio of production
fractions of $\Sigma$ over $\overline{\Sigma}$ in $\overline B$
decays.
\bibitem{dilution}
Gronau and Rosner, Ref.~\cite{gr}, discussed the measurement of the
dilution and of CP violation when
the charge of primary hadrons close by in phase space is used as
tags.
The
method is much more general, and can be applied to any conceivable
tag, as
we proceed to do here.  Even studies of $B_s - \overline B_s$ mixing
become feasible.
\end{references}
\end{document}